\documentstyle[pre,multicol,aps,epsf]{revtex}

\begin{document}
\draft

\title
{\Large \bf 
Phase synchronization in coupled nonidentical excitable systems and array
enhanced coherence resonance 
}

\author{ Bambi Hu$^{a,b}$, Changsong Zhou$^a$ }

\address 
{ $^a$ Department of Physics and Centre for Nonlinear Studies, 
Hong Kong Baptist University, Hong Kong, China\\ 
$^b$ Department of Physics, University of Houston, Texas 77204}

\maketitle

\begin{abstract}

We study the dynamics of a lattice of coupled
nonidentical Fitz Hugh-Nagumo system subject to independent external noise.   
It is shown that these stochastic oscillators can
lead to global synchronization behavior {\sl without an external signal}. 
With the increase of the noise intensity, the system exhibits coherence 
resonance behavior. Coupling  can  enhance greatly 
 the noise-induced coherence in the system.

\end{abstract}

\begin{multicols}{2}

The study of coupled oscillators is one of the fundamental 
problems  with applications in various fields~\cite{kur}. Mutual 
synchronization of the oscillators is of great interest and 
importance among the collective dynamics of the coupled 
oscillators. The notation of synchronization has been extended
to include a variety of phenomena in the context of interating 
chaotic oscillators, such as complete synchronization~\cite{fpp},
generalized synchronization~\cite{rul}, phase synchronization~\cite{rp1,opr}
 and lag synchronization~\cite{rp2}.  
Also, synchronization phenomenon has been studied in stochastic
systems. Stochastic resonance can be understood from the view 
point of frequency locking and phase synchronization of the noise-induced motion to the 
external signal~\cite{gms}.  Due to this synchronization, 
stochastic bistable elements subject to the 
{\sl same} periodic signal, when appropriately coupled, can display  global synchronization to 
the external periodic signal. This synchronization has the effect of  
enhancing  the stochastic resonance  in the array~\cite{lmd}.

Noise-induced coherent  motion has been a
topic of great interest recently. 
For a system at a  Hopf  bifurcation point,
an optimal amount of noise  can induce the most coherent 
motion in the system~\cite{hdnh}. More generally, noise-induced
coherent motion has been demonstrated in a variety of 
excitable systems~\cite{pk,l,lnk} and system with delay~\cite{os}. 
Although described by different terms such as 
stochastic resonance without  external periodic force, autonomous stochastic 
resonance or coherence resonance,  a common feature of this type of  systems is
the increased coherence of the motion and resonant like behavior of the 
coherence  induced {\sl purely by noise 
without an external  signal}.  An interesting question about this type of systems 
is how these stochastic elements behave when coupled. Do the
elements subjected to independent noise display synchronization?
and can the synchronization enhance the coherence in the system motion?
A recent study demonstrated  noise-induced global oscillation  and 
resonant behavior 
in a subexcitable media~\cite{hsg}. In Ref.~\cite{hyps}, 
it is shown that {\sl two} interacting 
coherence oscillators can be synchronized. However, whether the coherence
of the motion can been enhanced by the interaction between the elements are not
clear. 
 
This paper studies  these problems by the following $N$ 
 coupled {\sl nonidentical}
Fitz Hugh-Nagumo (FHN) neurons, a simple but representative model of 
excitable system and nerve pulses~\cite{pk}, 
\begin{eqnarray}
\epsilon\dot{x_i}&=&x_i-\frac{x_i^3}{3}-y_i+ g(x_{i+1}+x_{i-1}-2x_i),\\
\dot{y_i}&=&x_i+a_i+D\xi_i(t), 
\end{eqnarray}
where $a_i$ is a parameter of the $i$th  element. For a single FHN model,
if $|a|>1$, the system has only a stable fixed point, while $|a|<1$ a
limit cycle. The system with fixed point dynamics ($|a|$ slightly larger than one)
is excitable because
it  will return to the fixed point only after a large excursion (``near limit cycle'')
when perturbed away from the fixed point. To make the study more general,
we suppose $a_i$ is not the same for the elements, but has a uniform 
distribution $a_i\in (1, 1.1)$. This implementation of the model is 
physically significant because many physical systems  are diffusively coupled,
and nonidentity is more natural in physical situations.    
With nonidentical $a_i$, the  uncoupled elements 
will have different response to the same level of noise $\xi_i(t)$, e.g.,  different 
average firing frequency. The 
Gaussian noise is uncorrelated in different elements, i.e., 
$\langle \xi_i(t)\xi_j(t^{\prime})\rangle=\delta_{ij}\delta(t-t^{\prime})$.  
Periodic boundary
condition is employed  in our studies. The parameter of the time scale is 
fixed  at  $\epsilon=0.01$. 

To characterize synchronization behavior in the lattice of nonidentical 
excitable systems, we introduce the phase of the elements~\cite{gms}
\begin{equation}
\phi_i(t)=2\pi \frac{t-\tau_k}{\tau_{k+1}-\tau_k}+2\pi k,
\end{equation}
where $\tau_k$ is the time  of the $k$th firing of the element defined  in simulations 
by threshold crossing of $x_i(t)$ at $x=1.0$.
The quantity 
\begin{equation}
s_i=\sin^2\left(\frac{\phi_i-\phi_{i+1}}{2}\right)
\end{equation} 
measures the phase synchronization effect of neighbouring elements~\cite{opr}.
A spatiotemporal average of $s_i$, i.e., 
\begin{equation}
S=\lim\limits_{T\to \infty}\frac{1}{T}\int_0^T \left(\frac{1}{N}\sum\limits_{i=1}^{N}s_i\right) dt
\end{equation}
 gives a measure of the  degree of phase synchronization in
 the coupled system. For completely unsynchronized motion $S\approx 0.5$, while for globally synchronized system $S\approx 0$.

To measure the temporal coherence of the noise-induced  motion, 
we examine the distribution
of the pulse duration $T_k=\tau_{k+1}-\tau_k$. For a single element subject to 
noise, the distribution $P(T)$ has a peak at a certain value of $T_k$ and 
an  exponential tail 
at large values~\cite{pk}.  A measure of  the  sharpness of 
the distribution, for example,
\begin{equation}  
R=\langle T_k \rangle /\sqrt{\hbox{Var}(T_k)},
\end{equation}
which can be viewed as signal-to-noise ratio,  
provides  an indication of the coherence of the firing event. 
Biologically, this quantity is of  
importance  because it is related to the timing precision of the information
processing in neural systems~\cite{pwm}.  For a single element, 
it has been shown 
that $R$ possesses an optimal value at a certain level of noise~\cite{pk}.
 Here we adopt the same
measure of coherence,  with the distribution $P(T)$ constructed by pulse 
duration of all the $N$ elements during a long enough period of  time.  

Typical behaviors observed in numerical simulations of the system 
are in the following:

(1) Without external noise, each element comes to a fixed point state and no 
firing takes place. The fixed points are  slightly different due to the nonidentity.

(2) When strong enough noise is added to the originally quiescent system, the 
system is excited. The firing process of the elements may be in phase synchronization 
if the coupling is strong enough.

(3) The coherence of the temporal and spatial pattern of the firing process is 
greatly enhanced  in a region of noise and coupling strength.

The main features of the system are illustrated in 
Fig. 1, where  the results of $R$ and $S$ in the parameter 
space $(\log_{10}(g), \log_{10}(D))$ are shown for a lattice of $N=100$ elements.
For a fixed coupling $g$, $R$ increases first with the noise level $D$, reaches a 
maximum,  then decreases again, showing the typical resonant behavior {\sl without} 
an external signal. In general, for a stronger coupling, a higher level of noise is needed to excite
the system.  
Similarly, for a fixed noise level $D$, $R$ increases  with increasing $g$ until
it reaches an optimal value; after that, it decrease again.

From Figure 1, one can observe several dynamical regimes in the systems. 
For very
weak coupling ($g<10^{-2}$), the firing of the  elements are essentially independent, because 
a noise-induced 
firing of an element cannot excite its neighbouring lattices. Due to the 
independent firing, the phase difference has a uniform
random distribution on $(0, 2\pi)$, resulting $S\approx 0.5$. In this region, 
with the increase of $D$,
the temporal behavior of the system display coherence 
resonance similar to that observed in a single
element. The enhancement of the coherence by the noise  is not very pronounced. 
A maximal $R\approx 4.5$ is found 
around $D\approx 10^{-1}$.  A typical spatiotemporal pattern of $x_i$ and $s_i$ for weakly coupled
elements  subject to relatively weak noise is shown in Fig. 2(a).  Both the spatial and temporal
behaviors are quite irregular. Each element has its 
individual  average firing frequency 
due to the nonidentity (smaller frequency for larger $a_i$).

With the increase of the coupling strength, the system becomes sensitive to weak noise 
because the firing events induced by noise now  become the source of
 excitation 
of the neighbouring elements. This mutual excitation enhances the coherence of the motion 
in the coupled system, as  indicated by increasing $R$ and decreasing $S$. 
The lattice 
displays clusters of synchronization. The clusters breaks and reunites  during the evolution,
so that each element has slightly different firing frequency. Typical behavior of
this partial synchronization regime is shown in Fig. 2(b). 
However, if the 
external noise is strong enough so that the noise  dominates over the coupling, 
 firing  of each element is governed mainly by its individual noise, and  synchronization clusters 
cannot survive, as seen in Fig. 2(c). The coherence of the temporal
behavior is relatively low because quite large noise deforms greatly the ``near limit''
cycle. Note that firing frequency
is not locked, but the difference is not as pronounced as in  weak noise region (Fig. 2(a)).     
The next regime  where $R$  takes large values ($R\sim 18$) while $S$ is
very close to zero, is the most interesting, because the system performs quite regular motion 
globally, as seen in Fig. 2(d). All elements are locked to a relatively large firing frequency, and 
the  distribution of the pulse duration becomes very sharp. 
After that, with stronger coupling, the system keeps
global synchronization, however, the temporal behavior becomes  irregular again, as indicated by 
decreasing $R$. This can be
understood qualitatively from the global dynamics  $X=\langle x_i\rangle_N$ and 
 $Y=\langle Y_i\rangle_N$,  
with $\langle  \rangle_N$ denoting average over the lattice.  From Eqs. (1-2), we get approximately
\begin{eqnarray}
\epsilon \dot{X}&=&X-\frac{X^3}{3}-r^2X-Y,\\
\dot{Y}&=&X+a_0+\frac{D}{\sqrt{N}}\xi(t),
\end{eqnarray} 
where $r^2=\langle (x_i-X)^2\rangle_N$ is the fluctuation level  of local dynamics and higher order terms 
of this fluctuation  are ignored;  
$a_0=\langle a_i\rangle_N$. The summation of the independent noise is still a 
Gaussian noise $\xi(t)$, but with a weaker strength $D/\sqrt{N}$.
For strong enough  coupling, the system achieves global synchronization,  
$x_i\approx X$, and $r^2\approx 0$, and the coupled system
can be viewed  as a {\sl single} element  subject to a white noise with strength
$D/\sqrt{N}$. For a single element, a rather weak noise may not
excite the system, or the excited motion is  quite irregular~\cite{pk}. This explains the 
globally synchronized but irregular motion of the lattice, as shown in Fig. 2(e).   

The locations of the above five representative 
dynamical regimes shown in Fig. 2 have been indicated by the 
black dots in Fig. 1. These regimes are  typical for different size of the lattice. For 
a larger lattice, the regime of global regular motion ($R$ large and $S\sim 0$) is wider 
in the parameter space.     

Now let us discuss the phenomenon of {\sl array enhanced coherence resonance}. 
For a fixed size of lattice and a certain value of coupling $g$, there is an 
optimal level of noise at which the system has a maximal value $R_{max}$. 
Figure 3 shows  $R_{max}$ as a function of $g$ for different size of the lattice.
The dashed line represents  $R_{max}$ for  a single element with $a=1.05$. 
As seen from this
figure,  in the  weak coupling region,  $R_{max}$ of  the coupled lattice is lower than that  of a 
single element, because the nonidentity makes the distribution of the pulse  duration  broader.
However, in some intermediate  coupling region, even   
only {\sl two} coupled  elements can enhance the coherence of
 the temporal behavior, even though they are nonidentical. 
The enhancement is larger for larger $N$.  
For large enough lattice,  $R_{max}$ seems to be saturated, but 
the region of coupling in which $R_{max}$ takes large values is broader for larger lattice.   
As pointed 
above, for strong enough coupling, the coupled system tends to act as a single element, 
and $R_{max}$ converges to that of the single element at large  value of $g$. 
Clearly, for a larger lattice, stronger coupling is needed to  make the whole lattice
act as a single element due to the local diffusive coupling, resulting in  a slower convergence of  
$R_{max}$ to that of the  single element. 
 
The properties observed here have some  
similarity to  that in coupled bistable stochastic 
resonance oscillators {\sl subject 
to the same periodic signal}~\cite{lmd}.  
The difference is that in Refs.~\cite{lmd} and ~\cite{hsg}, 
global synchronization is observed only at an optimal level 
of noise, while in the present system, global synchronization 
occurs for strong enough coupling if the noise level is not too high.
The global motion can be regular or irregular.   
Here we should  emphasize that there is no periodic signal in our system, and the coherent motion
is {\sl purely induced by noise and enhanced by coupling}.   
      
To conclude, we demonstrate global synchronization and array enhanced coherence in a lattice
of locally coupled, nonidentical FHN model  neurons. The results are  similar for identical neurons. 
Interaction between the elements not only renders
synchronization of the firing process induced by independent noise, but also improves the 
coherence of the noise-induced motion greatly.  The phenomena  demonstrated may be of importance
in neurophysiology, where   background noise may play  an active role to  increase the order and 
timing precision  of a 
large ensemble of interacting neurons in biological information processing.

Zhou Thanks Prof. Gang Hu and Dr. Jinghua Xiao for helpful discussions.
This work is  supported in part by grants from the Hong Kong Research Grants Council (RGC) and the Hong Kong
Baptist University Faculty Research Grant (FRG).

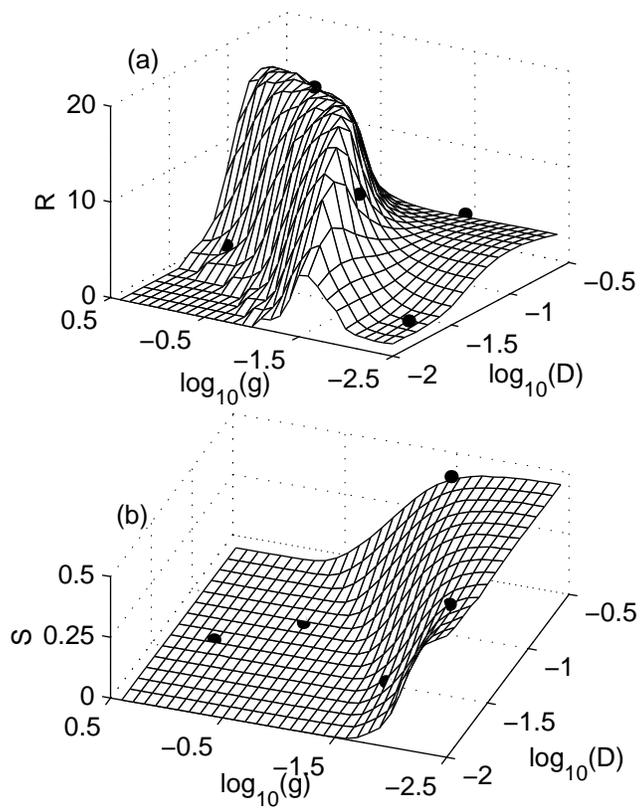
\begin{figure} 
\narrowtext
\caption{
Signal-to-noise ratio $R$ (a) and degree of phase synchronization $S$ (b) in the parameter space
$(\log_{10}(g), \log_{10}(D))$ of a coupled lattice of  nonidentical FHN neurons  with $N=100$.    
}
\end{figure} 

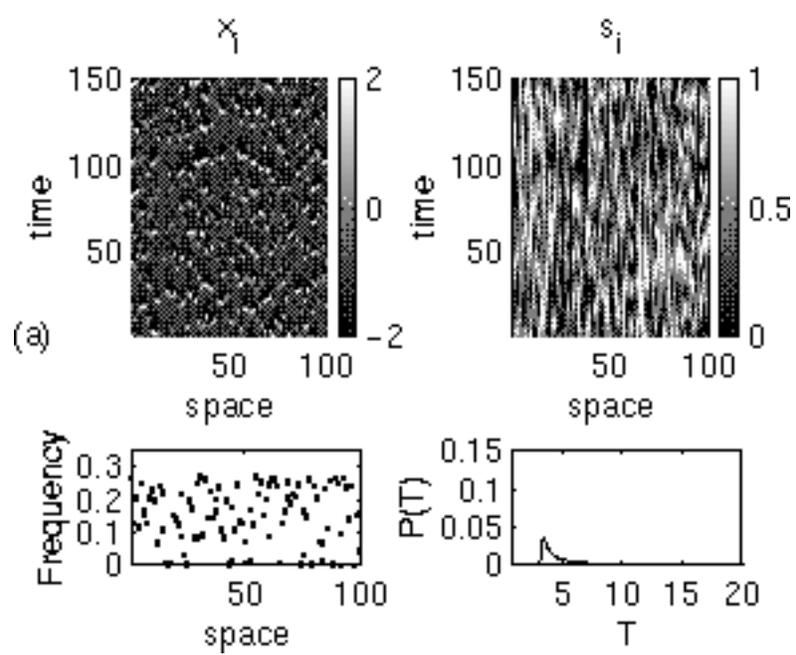
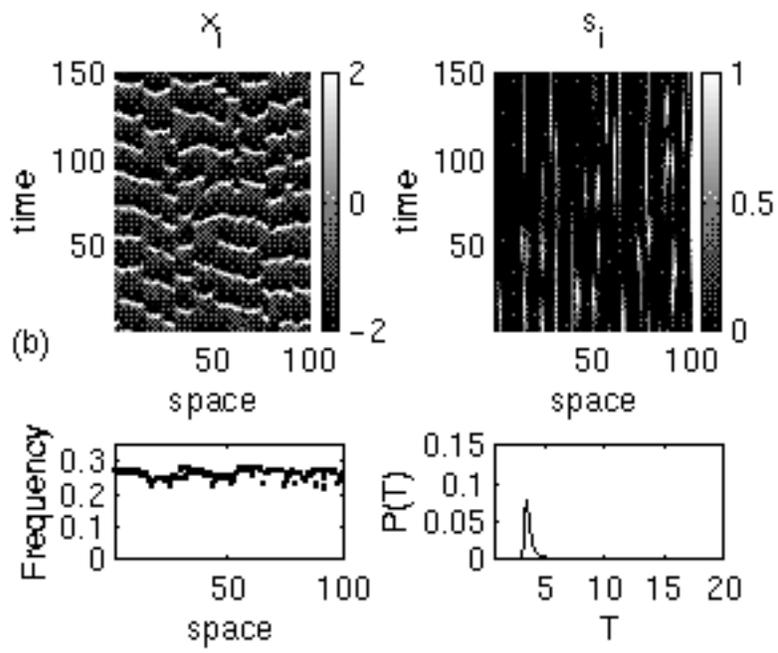
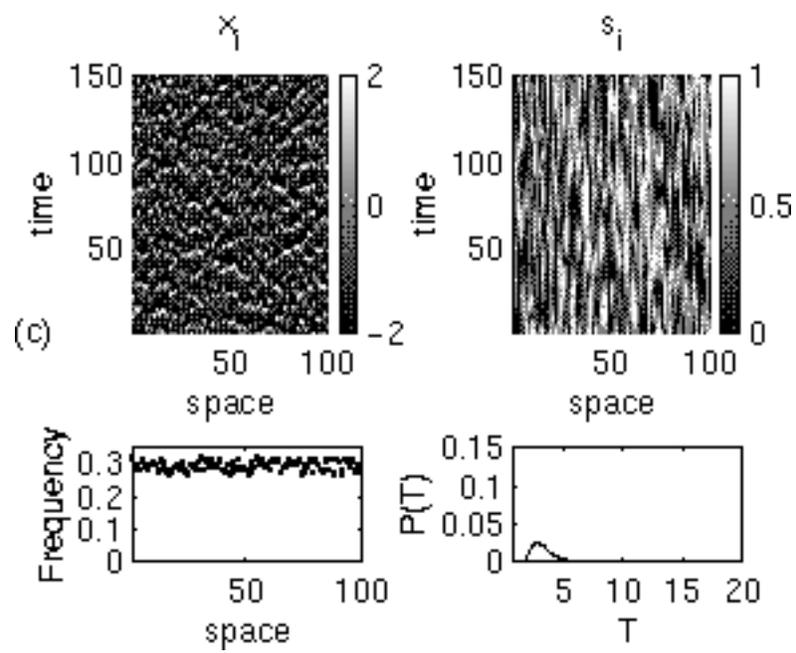
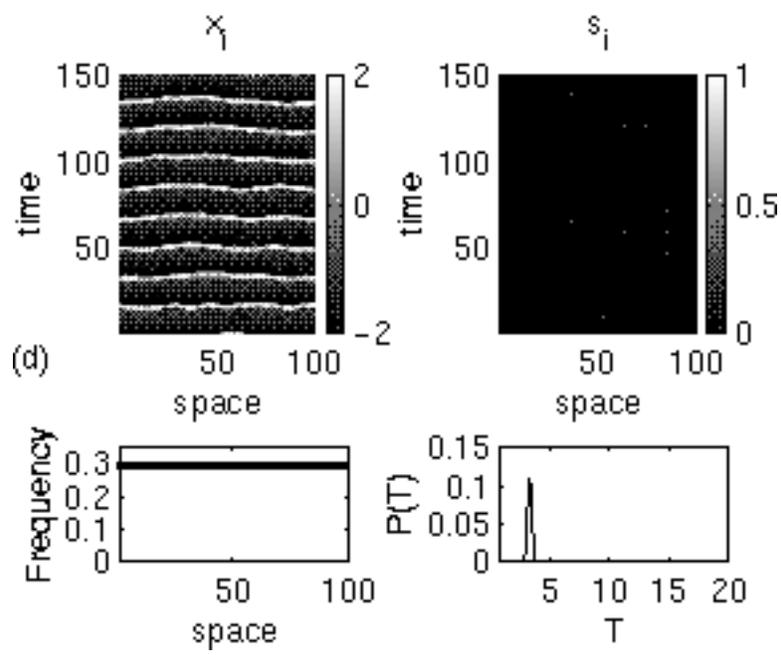
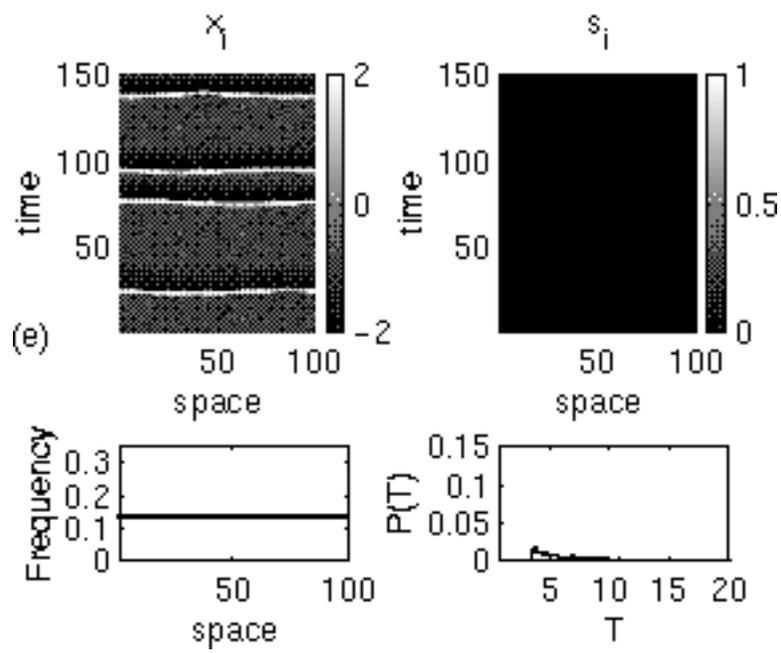
\begin{figure}
\narrowtext
\caption{
Five typical dynamical regimes in the parameter space. The upper panels show the spatiotemporal
structure of $x_i$ and $s_i$. The time step is 0.2.
  The lower panels show the average firing frequency of the lattice
and the distribution of the pulse duration. The location of these representative dynamical
behaviors are shown by black dots in  the parameters space in Fig. 1. The parameters are 
(a) $g=0.005, D=0.02$; (b) $g=0.03, D=0.025$; (c) $g=0.02, D=0.25$; (d) $g=0.25, D=0.07$; and (e) $g=1.0, D=0.04.$      
}
\end{figure}

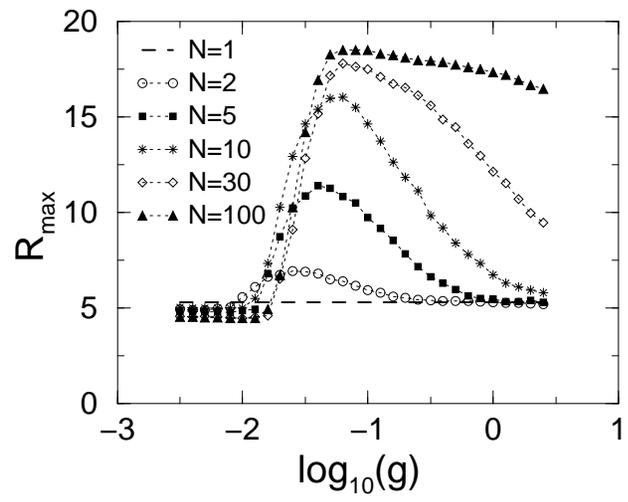
\begin{figure}
\narrowtext
\caption{
Illustration of array enhanced coherence resonance. The maximal value of $R$ is plotted as 
a function of $g$ for various  sizes of the lattice.
}
\end{figure}

\end{multicols}
\end{document}